\pdfoutput=1
\documentclass{article}
\usepackage{graphicx, indentfirst, hyperref, natbib}
\usepackage[a4paper, margin = 3cm]{geometry}
\makeatletter
\let\mailmark\@fnsymbol
\makeatother

\newcommand*{\cf}{\emph{cf.}}
\newcommand*{\eg}{\emph{eg.}}
\newcommand*{\etc}{\emph{etc}}
\newcommand*{\figref}[1]{Figure \ref{fig:#1}}
\newcommand*{\secref}[1]{Section \ref{sec:#1}}
\let\thxmark\textsuperscript
\let\cite\citep
\let\citeasnoun\citet
\begin{document}
\title{%
	\emph{Mamba}: a systematic software solution
	for beamline experiments at HEPS%
}
\author{%
	Yu Liu\thxmark{1,\mailmark{1}}, Yan-Da Geng\thxmark{2},
	Xiao-Xue Bi\thxmark{1}, Xiang Li\thxmark{1,3},\\
	Ye Tao\thxmark{1,3}, Jian-She Cao\thxmark{1,3},
	Yu-Hui Dong\thxmark{1,3}, Yi Zhang\thxmark{1,3,\mailmark{1}}%
}
\date{}
\maketitle
\begingroup
\renewcommand{\thefootnote}{\fnsymbol{footnote}}
\footnotetext[1]{\ %
	Correspondence e-mail:
	\texttt{liuyu91@ihep.ac.cn}, \texttt{zhangyi88@ihep.ac.cn}.%
}
\endgroup
\footnotetext[1]{\ %
	Institute of High Energy Physics, Chinese Academy of Sciences,
	Beijing 100049, People's Republic of China.%
}
\footnotetext[2]{\ %
	Kuang Yaming Honors School, Nanjing University,
	Nanjing 210093, People's Republic of China.%
}
\footnotetext[3]{\ %
	University of Chinese Academy of Sciences,
	Beijing 100049, People's Republic of China.%
}

\section*{Synopsis}

Keywords: Bluesky; experiment control; fly scan;
high-throughput experiment; software architecture.

\emph{Mamba} is a \emph{Bluesky}-based experiment control framework being
developed for HEPS; its frontend and backend collaborate through a RPC service,
and most importantly command injection.  Improvement of \emph{Bluesky}'s support
for high-frequency and high-throughput applications is in progress, with
\emph{Mamba Data Worker} as a key component.  Other plans, including an
experiment parameter generator and \emph{Mamba GUI Studio}, are also discussed.

\section*{Abstract}

To cater for the diverse experiment requirements at the High Energy Photon
Source (HEPS) with often limited human resources, \emph{Bluesky} is chosen as
the basis for our software framework, \emph{Mamba}.  In our attempt to address
\emph{Bluesky}'s lack of integrated GUIs, command injection with feedback is
chosen as the main way for the GUIs to cooperate with the CLI; a RPC service
is provided, which also covers functionalities unsuitable for command injection,
as well as pushing of status updates.  In order to fully support high-frequency
applications like fly scans, \emph{Bluesky}'s support for asynchronous control
is being improved; to support high-throughput experiments, \emph{Mamba Data
Worker} (\emph{MDW}) is being developed to cover the complexity in asynchronous
online data processing for these experiments.  To systematically simplify the
specification of metadata, scan parameters and data-processing graphs for each
type of experiments, an experiment parameter generator (EPG) will be developed;
experiment-specific modules to automate preparation steps will also be made.
The integration of off-the-shelf code in \emph{Mamba} for domain-specific
needs is under investigation, and \emph{Mamba GUI Studio} (\emph{MGS}) is
being developed to simplify the implementation and integration of GUIs.

\section{Introduction}

With the upgrade of synchrontron radiation facilities across the world,
great progress is being continously made in providing X-ray beams with better
emittance and coherence, as well as employing optical components and detectors
with higher performance.  Experiments with high communication frequencies or
high data throughputs, as well as experiments involving multiple modes, complex
\emph{in situ} environments or automated changing of samples, are becoming
increasingly prevalent.  While allowing for the multi-scale, multi-feature
and \emph{in situ} characterisation of samples, this also poses fundamental
challenges to experiment control and data acquisition/processing, both in
experiments themselves and the preparation steps before them.  The large number
of beamlines at many facilities, especially new facilities under construction,
also result in the hard demand to implement diverse experiment requirements
with a manageable codebase.  At the High Energy Photon Source (HEPS)
\cite{jiao2018}, a 4th-generation synchrotron radiation facility, where 14
beamlines will be provided in 2025 in its Phase I and up to 90 beamlines in
total can be served in further phases, all issues above are to be expected.
In order to address these issues while keeping our codebase maintainable
with often limited human resources, proper architecture design must
be carried out for the software components involved.

The foundation of \emph{Mamba}, our software framework, is the Python-based
\emph{Bluesky} \cite{allan2019}; before making the choice, we had researched
multiple well-known alternatives for similar applications, like \emph{GDA}
\cite{gibbons2011}, \emph{Sardana} \cite{coutinho2011}, \emph{Karabo}
\cite{hauf2019} and \emph{py4syn} \cite{slepicka2015}.  Here we avoid discussing
the details of our choice, and instead note that the choice is not based on the
availability of readily usable features, but based on the total efforts needed
to adapt the publicly available codebase to our applications.  When saying
``total efforts'', we not only include the efforts in development and
maintenance of our own codebase, but also include those in understanding, fixing
and customising the provided codebase.  After our research, we concluded that
because of the quite well-designed device interfaces (classes from \emph{ophyd})
in conjunction with the simple yet relatively powerful mechanism to combine them
in interlocked actions (\verb|RunEngine| from \emph{bluesky}) and represent
extracted data in a friendly format (the ``documents'') in real time,
\emph{Bluesky} is likely to fulfill a satisfactory fraction of the requirements
at HEPS with the best cost-to-effect ratio.  A first issue with \emph{Bluesky}
is the lack of integrated graphical user interfaces (GUIs); we discuss our
approach to this issue in \secref{backend}.  For other challenges we note
in the above, we give our plan of ongoing development in \secref{outlook}.

\section{\emph{Mamba}'s backend and frontend}\label{sec:backend}

\begin{figure}[htbp]\centering
\includegraphics[width = 0.96\textwidth]{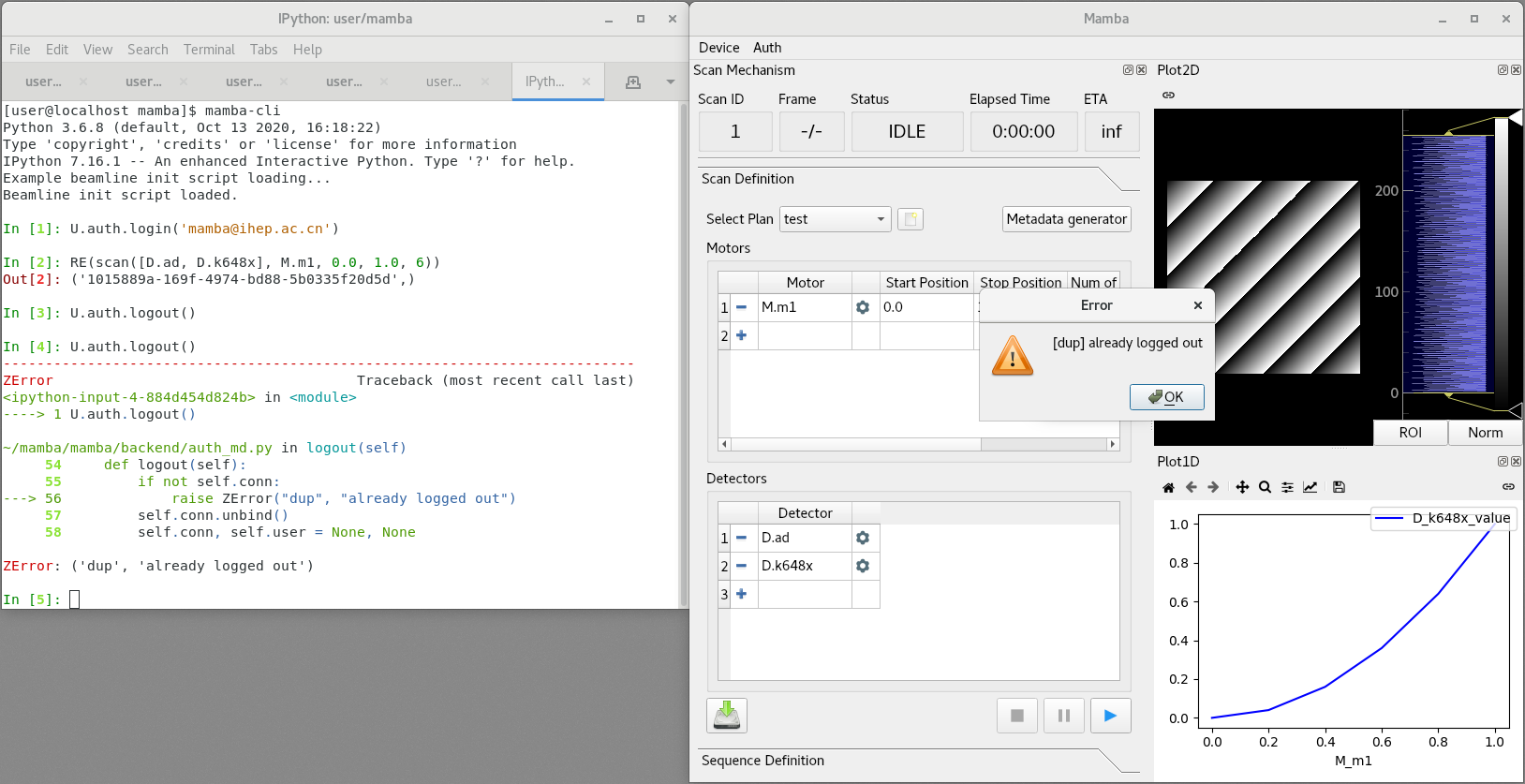}
\caption{%
	Command injection in \emph{Mamba}, showing an error caused by
	attempting to log out twice after a successful experiment session;
	we explicitly note that appearance of the GUIs may vary
	in the future due to ongoing frontend refactoring%
}\label{fig:cmd-inject}
\end{figure}

The first issue we observe with \emph{Bluesky}, in comparison with its easily
composable programming interfaces, is the lack of integrated GUIs.  For some
experiment tasks this is more like just an obstacle to users with a relatively
weak background in programming, but there are also tasks that are fundamentally
easier with GUIs than only with keyboards.  One good example is a requirement
from the hard X-ray high-resolution spectroscopy beamline (B5) at HEPS
\cite[\cf\ also][]{huotari2017}, where many regions of interest (ROIs) need to
be specified that properly cover individual light spots in sample images taken
from area detectors (and light spots in images that will follow); another
example is the manipulation of data-pipe graphs for \emph{Mamba Data Worker}
(\cf\ \secref{outlook}).  Since \emph{Bluesky} recommends the \emph{IPython}
interactive command line interface (CLI) of Python for its regular use on
beamlines, we have designed \emph{Mamba} with cooperation between the CLI
(\emph{Mamba} backend) and our GUIs (\emph{Mamba} frontend) in mind; inspired
by \emph{AutoCAD}-like software (called \emph{parametric modeling software}
in its industry), we extensively use what we call \emph{command injection}
(\figref{cmd-inject}) to implement this cooperation.  Before discussing the
communication architecture of \emph{Mamba} in detail, we note that its frontend
is still very immature in terms of internal structure and robustness; however,
the architecture between the backend and frontend, which is the subject
of this section, has been tested successfully in a real
tomography experiment on a testbed for HEPS.

With command injection, we basically treat GUIs as \emph{code generators}:
most user operations with GUIs are translated into equivalent commands that
get injected into the CLI, where they are actually executed.  This design is
beneficial in many ways, in terms of both user friendliness and architectural
soundness.  Users can naturally learn to use the CLI from using GUIs, and those
more proficient in programming may abstract repeated tasks into succinct yet
reusable CLI snippets.  Sometimes in order to perform some tasks that are not
yet implemented or simply inflexible with GUIs, developers may even ask users
to execute a few lines of code in the CLI; this is particularly meaningful
when considering that developing the GUI for an application is typically much
more complex than developing its CLI counterpart.  The emphasis of GUIs as
code generators also helps developers to naturally design optimal interfaces
when implementing requirements, which increases modularity and consequently
facilitates maintenance (especially automated testing).  To summarise the
above up, command injection allows the CLI and GUIs to complement each
other constructively, reducing workload for both users and developers.

\begin{figure}[htbp]\centering
\includegraphics[width = 0.49\textwidth]{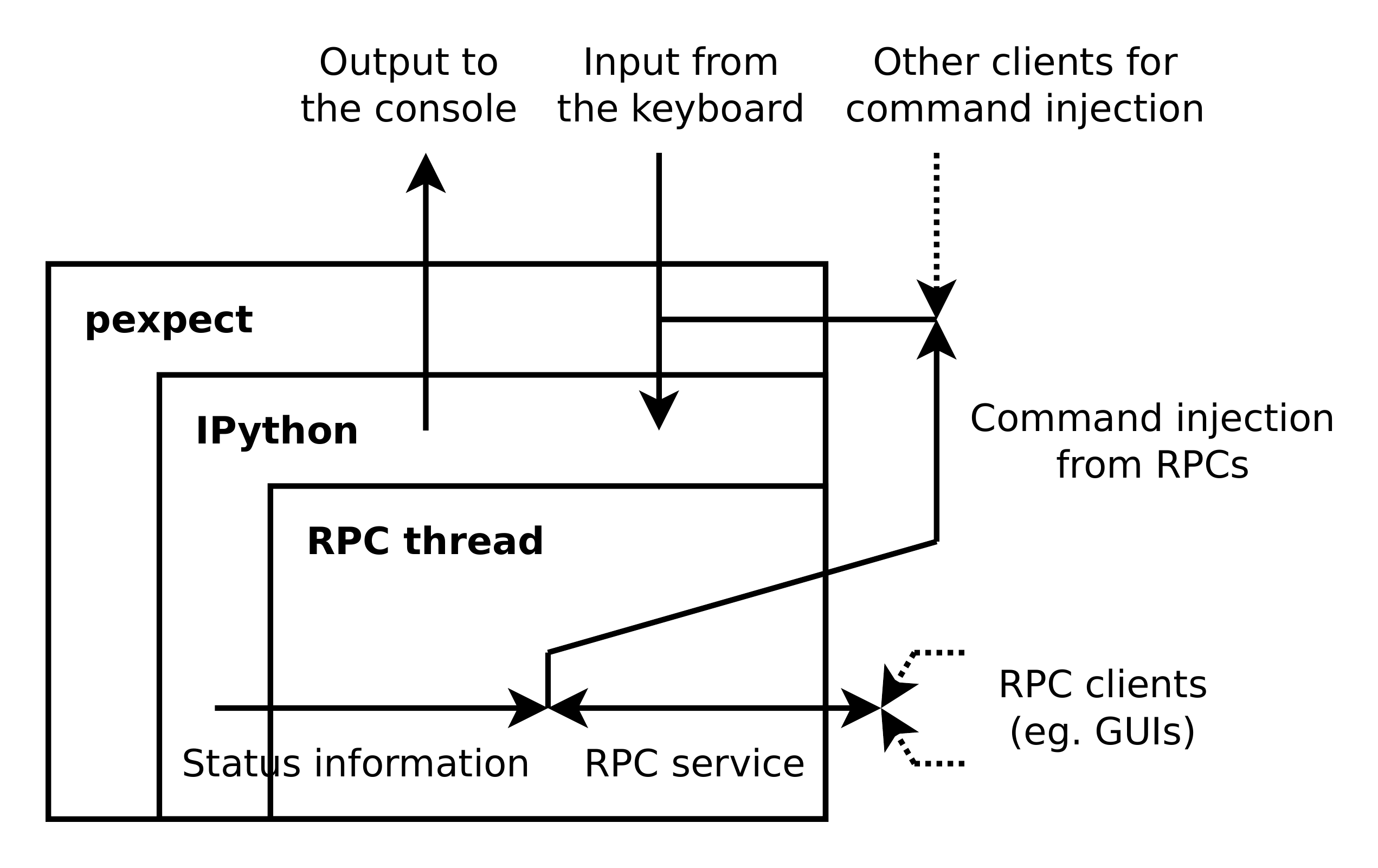}
\caption{\emph{Mamba}'s communication architecture}\label{fig:mamba-arch}
\end{figure}

\begin{figure}[htbp]\centering
\includegraphics[width = 0.75\textwidth]{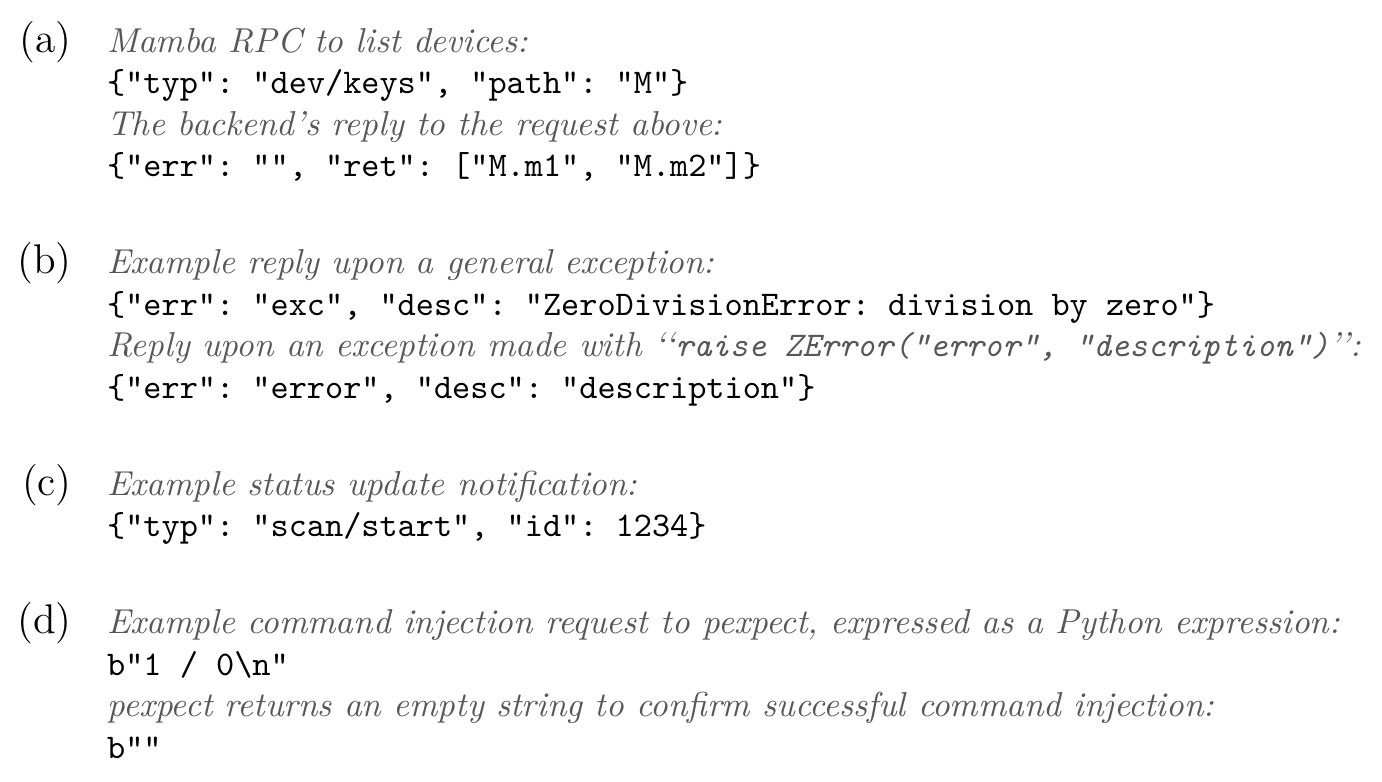}
\caption{%
	Example \emph{Mamba} communication: (a) normal request/reply,
	(b) replies upon errors, (c) notification, (d) raw communication
	to \emph{pexpect}; except for (d) which uses raw byte strings,
	all other (RPC) communication uses a JSON-based format%
}\label{fig:mamba-rpc}
\end{figure}

\begin{figure}[htbp]\centering
\includegraphics[width = 0.49\textwidth]{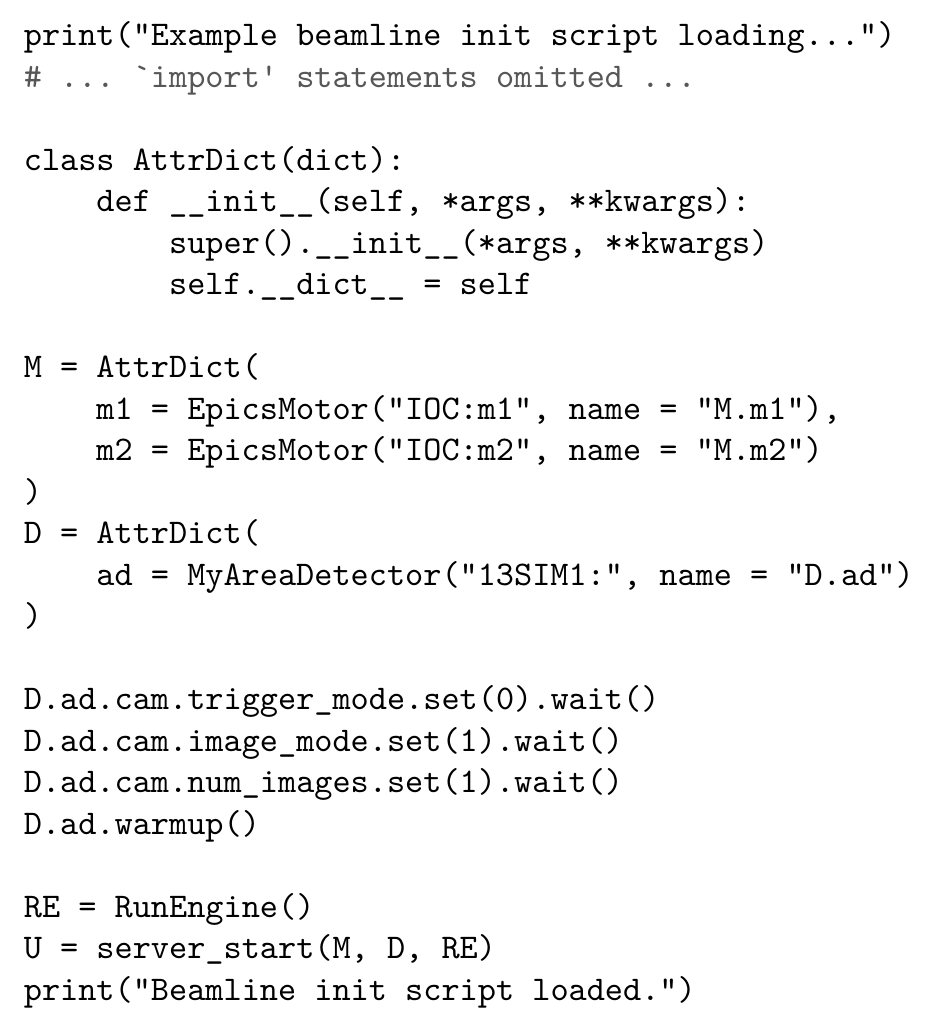}
\caption{Example \emph{IPython} startup script for \emph{Mamba}}
\label{fig:init-script}
\end{figure}

The actual communication architecture between the backend and frontend
of \emph{Mamba} is shown in \figref{mamba-arch}.  The backend is run as a
subprocess of a wrapping program, which is based on the \emph{pexpect} library
and forwards input from the user and output from the subprocess; the forwarding
program also listens on a socket (\emph{ZeroMQ} \verb|REP|) that \emph{command
injection clients} can connect to, which send the actual injection requests
(\figref{mamba-rpc}(d)).  This way, commands are injected as if they were from
the user's keyboard.  A problem with \emph{pexpect}-based command injection is
the lack of feedback: because the wrapping program does not understand the input
semantics of the program (\eg\ \emph{IPython}) it wraps, it only knows whether
some command has been successfully injected, instead of the final result (return
value or exception in Python, \cf\ \figref{cmd-inject}) of its execution.
For this reason, we encapsulate command injection with a remote procedure
call (RPC) service started by the \verb|server_start| function in the
\emph{Mamba}-specific startup script for \emph{IPython} (\figref{init-script});
the RPC service is a native Python thread with access to relevant
\emph{IPython} interfaces, so it can capture the results
of injected commands and return them to its clients.

So the \emph{Mamba} backend provides a RPC service that supports command
injection with feedback; however, there are also communication requirements
between the backend and frontend that are unsuitable for command injection.
A first example is passwords which should not appear in clear text on the CLI
because of the command line history mechanism; additionally, many status queries
(\eg\ listing known motors and detectors) are only required by GUIs, and are
inessential for CLI-only use of \emph{Mamba}, so the appearance of relevant
commands on the CLI would be mostly useless to users.  Therefore we also provide
\emph{special RPCs} for these requirements (\figref{mamba-rpc}(a)); however
because RPCs need dedicated encapsulation code (RPC-specific syntax checks
\etc), we have formed the policy that \emph{special RPCs should usually not
be added for communication essential for CLI-only use}.  Noticing the need for
the frontend to get status updates (\figref{mamba-rpc}(c)) and the intrinsic
weakness of polling (polling too often wastes system resources, while polling
too scarcely risks loss of updates), in addition to a \emph{request/reply
socket} (\emph{ZeroMQ} \verb|REP|) for regular RPCs, a \emph{notification
socket} (\emph{ZeroMQ} \verb|PUB|) is also provided by the RPC service
to proactively push updates to clients.

To further simplify our codebase, the finer details in \emph{Mamba} have
also undergone careful design, for which we give two examples.  One is the
introduction of \verb|ZError|, a subclass of Python's \verb|Exception|,
that can be raised by handlers in the RPC service to give fine-grained
reports for errors that have been anticipated by developers (\cf\ Figure
\ref{fig:cmd-inject} and \ref{fig:mamba-rpc}(b)).  The RPC service will return
information extracted from \verb|ZError| to clients, instead of the more generic
information it will return upon other types of exceptions; this way, RPC clients
can set up exception handlers accordingly, and only use a generic handler
as a last resort.  Another example is the use of variables \verb|M| and
\verb|D| to group motors and detectors respectively in the global scope
of \emph{IPython} (\cf\ Figure \ref{fig:cmd-inject}, \ref{fig:init-script}
and \ref{fig:mamba-rpc}(a)), just like how \verb|RE| is recommended by
\emph{Bluesky} developers for the \verb|RunEngine| instance in the same scope.
We find this much more natural as a way to mark the ``movability'' of devices
than alternatives, like passing two \emph{ad hoc} dictionaries as arguments to
\verb|server_start| which has the disadvantage that the device lists cannot
be modified dynamically.  We have even gone one step further and modified core
\emph{ophyd} code to allow device object names like \verb|M.m1| that contain
dots, so that we can enforce a policy that \emph{the name of a device object
must be a Python expression referencing exactly the same object},
which has proven to again save quite a lot of code.

\section{Further plans on \emph{Mamba}}\label{sec:outlook}

After a quite extensive research on the eligibility of \emph{Bluesky} for the
applications at HEPS, we concluded that apart from the lack of GUI integration,
\emph{Bluesky} is able to cover most low-frequency low-throughput (typically
with $<10$\,Hz communication between the computer and devices involved, and data
rates $<100$\,MB/s) needs at HEPS, in both regular step-scan experiments and
certain preparation steps (\eg\ the automated arming of samples, including the
fine tuning of their positions; \cf\ also the requirement from the B5 beamline
at HEPS mentioned in \secref{backend}).  Nevertheless, because HEPS is a
4th-generation synchrotron radiation facility, its small X-ray spots and high
brightness not only facilitates high-resolution imaging, but also necessitates
continuous scans (\emph{fly scans}) to handle the significantly larger number
of data points and the much more serious radiation damage of samples.  So
both high-frequency and high-throughput applications -- exactly where we
find \emph{Bluesky} to be currently not very good at -- are essential to
HEPS; if these weaknesses are somehow addressed, \emph{Bluesky} will be
able to provide a solid unified basis for beamline experiments at HEPS.

We begin with high-frequency applications, represented by fly scans; a
typical high-frequency (but low-throughput) application is sound recording,
which we compare fly scans to.  When performing fly scans (sound recording),
because regular computers cannot handle the influx of data points at too high
frequencies, we instead use dedicated controllers (sound recorders) to do the
handling; the computer reads data from the controllers in a block-by-block
(instead of point-by-point) fashion, and other than that just sends control
messages like ``start'', ``stop'' or ``pause''.  From the above we can see
that \emph{the key to high-frequency applications is asynchronous control}
(indirectly with dedicated controllers), which currently does not seem quite
easy to do with \emph{Bluesky}'s \verb|RunEngine|.  In fact the latter already
has primitive support for simple fly scans which only need ``start'' (the
\verb|kickoff| operation in \verb|RunEngine|) and ``stop'' (the \verb|complete|
operation), where the data readout is mainly done offline (the \verb|collect|
operation run after \verb|complete|).  At HEPS, we are currently exploring an
implementation of fly scans that allows \emph{real-time tuning of the scanning
behaviours} (adaptive speed/step-size tuning, automatic pausing/resuming \etc)
based on online processing of the data read from controllers.  This might
be of particular interest in requirements like obtaining the optimal X-ray
spot size and wavefront for focus alignment, as well as enabling ultra-high
stability of the sample and X-ray probe during multi-dimensional scanning
measurements at the hard X-ray nanoprobe beamline (B2) at HEPS.

\begin{figure}[htbp]\centering
\includegraphics[width = 0.64\textwidth]{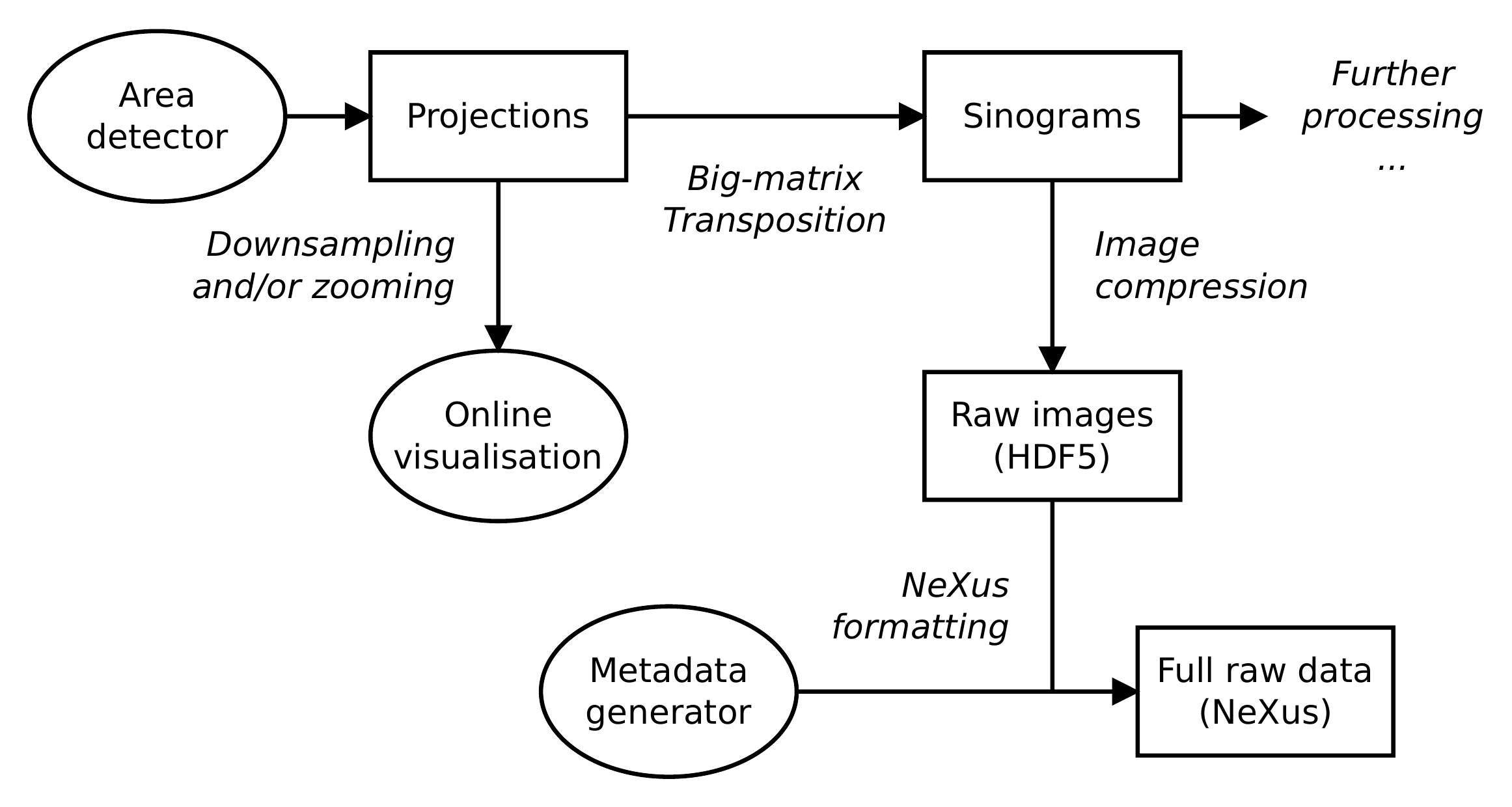}
\caption{%
	Partial data-pipe graph for a simple high-throughput tomography experiment%
}\label{fig:pipe-graph}
\end{figure}

\emph{Bluesky} supports online data processing using Python packages from
the \emph{SciPy} ecosystem, but this is currently done through synchronous
point-by-point callbacks; similar to how point-by-point processing of sound
data cannot be done synchronously with regular computers, synchronous processing
is unsuitable for high-throughput applications.  The solution is also similar --
use asynchronous processing instead; noticing the discussion in the previous
paragraph, we can see that the same asynchronous mechanism we envision also
covers the data-processing needs for high-frequency experiments naturally.
To address the complexity in both the asynchronous collaboration of worker
processes (buffering, polling/pushing, error handling \etc, plus the resource
management for processing of big data) and the diverse domain-specific logics
we need to support, we are developing what we call the \emph{Mamba Data
Worker} (\emph{MDW}) framework, which will be to an extent like \emph{HiDRA}
\cite{fischer2017} and \emph{Odin} \cite{yendell2017}.  However, instead of
focusing more on data producers, \emph{MDW} will treat producers and consumers
equally, also handling the diverse formating requirements of raw data from
beamline experiments at HEPS \cite{hu2021a}.  The same requirements are to our
knowledge not something actively pursued by \emph{Bluesky}'s \emph{databroker},
aside from the problem we notice that \emph{databroker} is not performant
enough when there is heavy disk I/O on the same machine.  Moreover, instead
of mainly supporting linear processing pipelines, \emph{MDW} will support
\emph{full-fledged graphs of data pipes} (\figref{pipe-graph}), perhaps
implemented in cooperation with the \emph{Daisy} project \cite{hu2021b};
this is crucial for the real-time tuning of fly scans, and will
also be very helpful in complex multimodal experiments.

As has also been discussed by \citeasnoun{hu2021a}, aside from the diverse
needs for processing of ``real'' data from experiments, the challenges posed by
the many types of experiments possible at HEPS also include the complexity in
management of the scientific metadata for these experiments, which will directly
affect the formatting of raw data by \emph{MDW}.  From the data producers' side,
for each type of experiment at a certain beamline, the number of parameters
intended to be tuned by users (especially typical users) are usually much
smaller than the total number of parameters for devices accessible on
the beamline.  Therefore it is also necessary to extend \emph{Bluesky}'s
\verb|RunEngine| for each type of beamline experiments, so that users' need
to care about irrelevant parameters are systematically minimised.  Noticing
the strong correlation between the specification of scientific metadata, device
parameters and data-pipe graphs, we are designing an \emph{experiment parameter
generator} (EPG) mechanism.  Given an \emph{experiment schema}, the EPG should
accept a minimised group of inputs; while the inputs are selected for typical
needs, the output should be in a form both customisable by advanced users
and friendly to automation mechanisms.  In the same spirit of simplifying
user operations, we will also develop \emph{Mamba} modules to automate
experiment-specific preparation steps as are mentioned in the beginning of
this section; considering the small X-ray spots allowed at HEPS, this will help
greatly in reducing time costed by tasks like the the fine tuning of beams.
We find that ``intelligent'' techniques, \eg\ those based on statistical
learning, are often useful in these steps, and we believe the architecture
of \emph{Mamba} will help to incorporate these techniques into our workflow.

For both the backend and frontend of \emph{Mamba}, we fully realise the
necessity to reuse off-the-shelf code from open-source projects to avoid
unnecessary duplication of efforts in providing domain-specific abstractions.
Integration of code from projects, \eg\ \emph{xrayutilities} and \emph{diffcalc}
(for \emph{SPEC}-like access to crystallographic coordinates) on the backend
side, as well as \emph{TomoPy} and \emph{PyFAI} (for requirements in tomography)
on the frontend side, are already under investigation.  We also realise the
dominant status of certain projects in specific fields, \eg\ \emph{MXCuBE}
\cite{oscarsson2019} in macromolecular crystallography, and will consider ways
to provide a smooth experience at HEPS to users familiar with these projects.
We may reuse \emph{Mamba} components just enough to integrate the upstream
project into the workflow at HEPS, or conversely integrate upstream components
into \emph{Mamba}, or even (if preferable) reimplement functionalities in
\emph{Mamba} and just emulate the GUIs for them; the actual way will be chosen
depending on the specific project, in friendly cooperation with upstream
developers, with the goal of minimising the efforts on both sides in mind.
We also note the necessity to support control systems other than EPICS, which
is certainly doable with \emph{Bluesky} but just not a focus to its developers.
This is imperative for high-throughput area detectors which are non-trivial to
support cleanly with the \emph{areaDetector} framework in EPICS, and we are
working on direct \emph{ophyd} support for them with \emph{MDW} integration.
There may also be systematic demand for other devices unsupported
by EPICS that cannot be replaced with supported workalikes,
which has fortunately not yet been encountered by us.

\begin{figure}[htbp]\centering
\includegraphics[width = 0.49\textwidth]{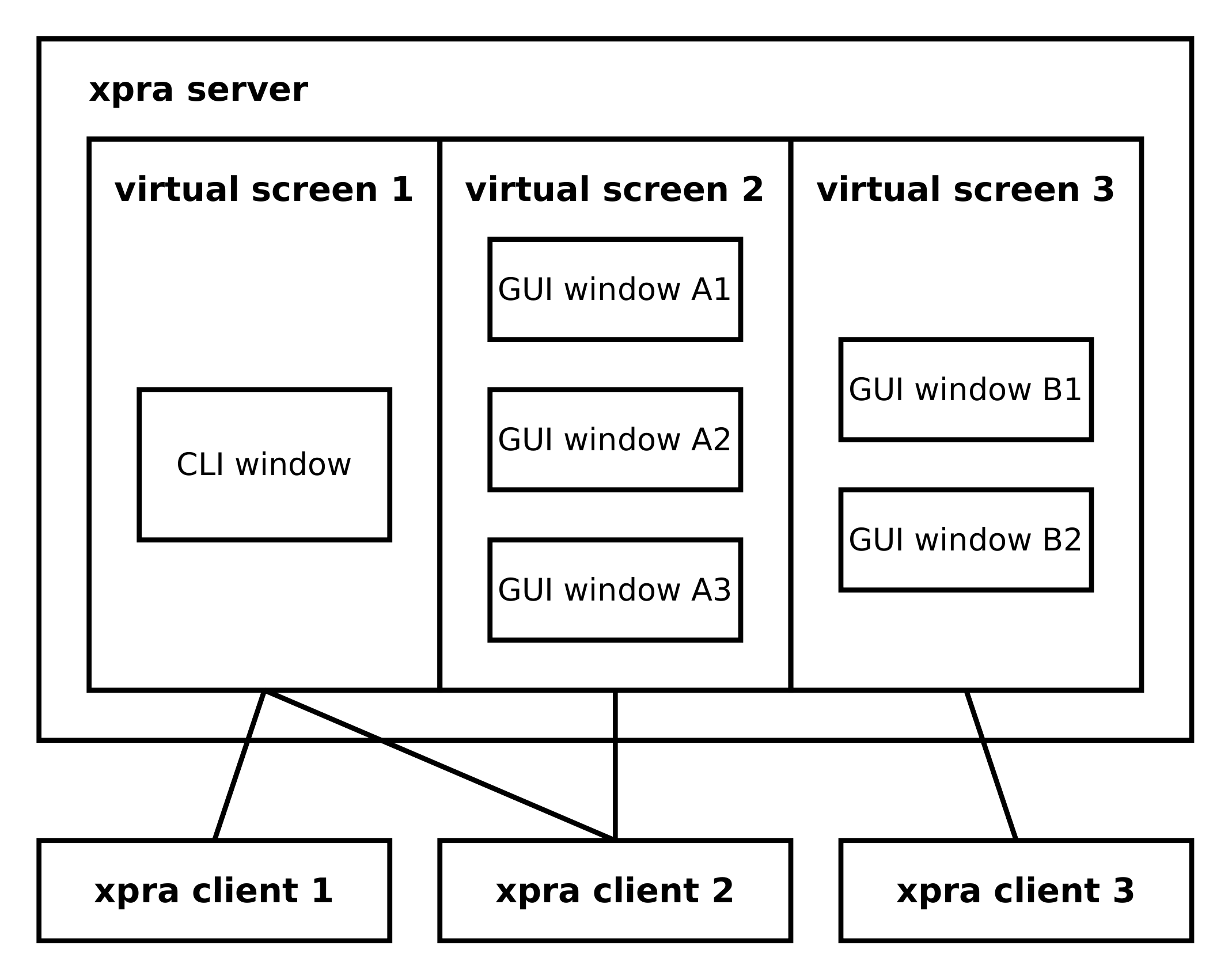}
\caption{%
	\emph{xpra} provides ``virtual screens'', each of which
	can be accessed by multiple clients simultaneously%
}\label{fig:xpra-tmux}
\end{figure}

To reduce the efforts necessary for implementation and integration of GUIs,
we are making what we call \emph{Mamba GUI Studio} (\emph{MGS}) to provide
reusable utility widgets and allow drag-and-drop composition of high-level GUI
components, similar to what is done by \citeasnoun{sobhani2020}.  A feature
commonly requested for \emph{Mamba} is cross-platform use of its frontend, but
we find it really complex to expose its RPC service (especially the command
injection mechanism) to the network without harming security.  For this reason,
we only allow the backend and frontend of \emph{Mamba} to run on the same host,
but meanwhile we plan to use the \emph{xpra} utility (\figref{xpra-tmux}),
with communication secured with SSH, to provide access on remote computers with
operating systems supported by \emph{xpra}.  Since \emph{xpra} supports multiple
coexisting ``virtual screens'' and simultaneous access to one virtual screen by
multiple clients, users can additionally collaborate either using self-chosen
groups of GUIs or using GUI groups shared by others; proper coordination is
obviously needed between users, perhaps using walkie-talkies or some
chat software, to avoid conflicts between their operations.

\section{Conclusion}

We are developing the \emph{Bluesky}-based \emph{Mamba} software framework for
the diverse experiment requirements at HEPS.  We use command injection with
feedback to allow the CLI and GUIs of \emph{Mamba} to complement each other
constructively; considering certain functionalities unsuitable for command
injection, \emph{Mamba} provides a RPC service which also supports proactive
pushing of status updates.  We take multiple measures to further simplify
the codebase of \emph{Mamba}, like the use of \verb|ZError| to simplify error
handling, and the use of \verb|M| and \verb|D| to respectively group motors
and detectors.  We find \emph{Bluesky}'s weaknesses in high-frequency and
high-throughput experiments to be exactly where it lacks in requirements at
HEPS, and plan to address the former by improving \emph{Bluesky}'s support
for asynchronous control.  To fully address the complexity in data processing
for high-throughput experiments, we are developing \emph{Mamba Data Worker},
which will cover the entire process from producers to consumers, as well as
support full-fledged graphs of data pipes.  To simplify the specification
of scientific metadata, device parameters and data-pipe graphs, we will
develop an experiment parameter generator; the generation will be tailored
systematically according to the experiment type, and its output will be
customisable yet machine-friendly.  Experiment-specific modules to automate
preparation steps will be developed similarly.  We are investigating the
integration of off-the-shelf code into \emph{Mamba} to provide domain-specific
functionalities, and are also developing \emph{Mamba GUI Studio} to
simplify the implementation and integration of GUIs.  We plan to use a
\emph{xpra}-based mechanism to allow cross-platform access to \emph{Mamba},
which will also enable easy collaboration between users.

\subsection*{Acknowledgements}

This work was supported by the National Basic Research Program of China
(2017YFA0504900), the One-Hundred Talents Program of Chinese Academy of
Sciences (Y851552), the Strategic Priority Research Program of Chinese
Academy of Sciences (XDB37000000), the Specialised Program for
Informatisation of Chinese Academy of Sciences (CAS-WX2021PY-0106)
and the National Science Foundation for Young Scientists
of China (Grant No.\ 12005253).

\bibliography{art4}
\end{document}